\documentclass[fleqn,usenatbib]{mnras}

\usepackage{newtxtext,newtxmath}

\usepackage[T1]{fontenc}
\usepackage{ae,aecompl}

\usepackage{graphicx}	
\usepackage{subfigure}
\usepackage{caption}
\usepackage[utf8]{inputenc}
\usepackage{amsmath}	
\usepackage{amssymb}	
\usepackage{soul}
\usepackage[dvipsnames]{xcolor}

\newcommand{\kms}{\mathrm{km~s^{-1}}}
\usepackage{etoolbox}
\makeatletter
\patchcmd\@combinedblfloats{\box\@outputbox}{\unvbox\@outputbox}{}{%
   \errmessage{\noexpand\@combinedblfloats could not be patched}%
}%
 \makeatother

\title[C. exchange \& rad-pressure on stellar/planet winds]{Hydrodynamical interaction of stellar and planetary winds: effects of charge exchange and radiation pressure on the observed Ly$\alpha$ absorption}

\author[Esquivel et al.]{
A. Esquivel,$^{1,2}$\thanks{E-mail:esquivel@nucleares.unam.mx}
M. Schneiter,$^{2,3,4}$
C. Villarreal D'Angelo,$^{5}$
M. A. Sgr{\'o}$^{2}$\newauthor
and L. Krapp,$^{2,6}$
\\
$^{1}$Instituto de Ciencias Nucleares, Universidad Nacional Aut\'onoma
de M\'exico, Apartado Postal 70-543, 04510 Ciudad de M\'exico, Mexico\\
$^{2}$Instituto de Astronom\'ia Te\'orica y Experimental, CONICET - UNC, Laprida 854, X5000BGR C\'ordoba, Argentina\\
$^{3}$Departamento de Materiales y Tecnolog\'ia, Universidad Nacional de C\'ordoba, X5016GCA, C\'ordoba, Argentina\\
$^{4}$Departament of Astronomy, AlbaNova, Stockholm University, SE-106 91 Stockholm, Sweden \\
$^{5}$School of Physics, Trinity College Dublin, College Green, Dublin 2, Ireland\\
$^{6}$Niels Bohr International Academy, The Niels Bohr Institute, Blegdamsvej 17, DK-2100, Copenhagen Ø, Denmark\\}
\date{Accepted. Received; in original form ZZZ}

\pubyear{2019}

\begin{document}
\label{firstpage}
\pagerange{\pageref{firstpage}--\pageref{lastpage}}
\maketitle
\begin{abstract}
Lyman $\alpha$ observations of the transiting exoplanet HD\,209458b enable the study of exoplanets exospheres exposed to stellar EUV fluxes, as well as the interacting stellar wind properties. In this study we present 3D hydrodynamical models for the stellar-planetary wind interaction including radiation pressure and charge exchange, together with photoionization, recombination and collisional ionization processes. Our models explore the contribution of the radiation pressure and charge exchange on the Ly$\alpha$ absorption profile in a hydrodynamical framework, and for a single set of stellar wind parameters appropriate for HD\,209458. We find that most of the absorption is produced by the material from the planet, with a secondary contribution of neutralized stellar ions by charge exchange. At the same time, the hydrodynamic shock heats up the planetary material, resulting in a broad thermal profile. Meanwhile, the radiation pressure yielded a small velocity shift of the absorbing material.
While neither charge exchange nor radiation pressure provide enough neutrals at the velocity needed to explain the observations at $-100~\kms$ individually, we find that the two effects combined with the broad thermal profile are able to explain the observations.
\end{abstract}




\section{Introduction}

After the first Ly$\alpha$ transit observations of the hot Jupiter HD 209458b were published \citep{vidal2003}, different scenarios with an increasing level of complexity have been proposed to explain them. In particular, the $\sim 10\%$ of absorption during transit at $100~\kms$ from the line centre towards the blue, has been used as a key feature when modelling these observations.

The blue-shifted Ly$\alpha$ absorption requires a population of neutral hydrogen at (or near) those velocities. Naturally, there are at least two ways to produce this population in a two-wind interaction model:
(i) to accelerate the neutrals injected by the planet (by radiation pressure for instance), and (ii) to neutralize already fast moving ions from the stellar wind (charge-exchange).

The first model developed to explain the presence of high velocity H neutrals can be found in the work of \citet{vidal2003}, who proposed that the stellar radiation pressure by itself is strong enough to accelerate them up to the observed velocities, reproducing the Ly$\alpha$ observations. This scenario was further developed in \citet*{bourrier2013} with a 3D particle/kinetic model where the neutral atoms are subject to radiation pressure as in \citet{vidal2003}, but also included ionization and self-shielding. In this work, the authors concluded that the radiation pressure alone is able to explain the Ly$\alpha$ observations when an ionizing flux of $3$ to $4$ times the solar value was considered together with a hydrogen escape rate from the planet in the range of $10^9$ and $10^{11}\,\mathrm{g\,s^{-1}}$.
In a similar work, \citet{kislyakova2014} modelled the Ly$\alpha$ transit observation including all the processes mentioned earlier (radiation pressure, natural broadening and charge exchange) by means of a Monte Carlo simulation, and estimated the stellar wind velocity and the planetary magnetic moment.

An alternative explanation was given by \citet{holmstrom2008} and \citet{ekenback2010} who also by means of a particle model proposed that the absorption at high velocities in the blue part of the line is due to energetic neutral atoms (ENAs), which result from the charge exchange between fast stellar wind ions and slow planetary H neutrals.

The presence of such ENAs to model the spectral features of the observed Ly$\alpha$ profile was also used in \citet{tremblin2013}, although in that study the authors used a 2D hydrodynamic model.
In their work, they employed a chemistry module to calculate the amount of hot neutral atoms that are produced by charge exchange, and found that the number of hot neutrals is sufficient to explain the transit Ly$\alpha$ observations. An improvement of this model can be found in \citet{christie2016},  who calculate the ENA production in a more realistic 2D axisymmetric model; and in \citet{shaikhislamov2016}, who takes into account the electron impact ionization as well as the photoionization and the dielectronic recombination as radiative processes together with the action of tidal forces.

Natural and thermal broadening have also been proposed as an explanation for the velocity structure in the Ly$\alpha$ line \citep{benjaffel2010,koskinen2010,koskinen2013}. \citet{koskinen2013} employed a 1D hydrodynamic model to explain that a single layer of H, in the thermosphere of the planet, is capable to reproduce the transit observation. Their results are in agreement with the empirical model presented in \citet{koskinen2010}.

\citet{schneiter2007} modeled the planetary mass loss rate, $\dot{M}_\text{p}$, with a 3D hydrodynamic simulation, estimating an upper limit for $\dot{M}_\text{p}$ within the hydrodynamic approximation. \citet{villarreal2014} further explored this model, with a variety of stellar wind conditions, as well as asymmetries for the planetary escaping exosphere. In these works, the ionization processes were included through an ionization temperature, and the radiation pressure was implemented as a uniform reduction of the stellar gravitational force. Both works where able to reproduce the observed Ly$\alpha$ profile within the parameter ranges explored in the numerical models. An important improvement to these models was presented in \citet{schneiter}, where the radiation processes, such as the photoionization of neutrals, collisional ionization and recombination were taken into account self-consistently with a ray-tracing approach, which yielded a better reproduction of the blue part of the Ly$\alpha$ line.

Most of the mentioned works have been able to either  predict or roughly reproduce the observed Ly$\alpha$ absorption (mainly in the blue), with different assumptions, either hydrodynamic or kinetic simulations.

The present work is an effort to study the Ly$\alpha$ emission that would result from hydrodynamical simulations similar to those presented in \citet{schneiter2007,villarreal2014,schneiter}, but including the effects of charge exchange as proposed by \citet{tremblin2013}, as well as the radiation pressure in a similar way to \citet{bourrier2013}.
We study the two processes separately and together to distinguish their relative relevance in the absorption seen in Ly$\alpha$.

Section \ref{model} presents the full set of hydrodynamic equations solved (\ref{sec:code}), and discusses the source terms (\ref{source}). Later, in \ref{parameters} the used stellar and planetary parameters are laid out.
The results and a discussion are presented in sections \ref{results} and \ref{discussion}, respectively. We present a summary with our conclusions in Section \ref{sec:conclusions}.
In appendix \ref{sec:ttA} we compare our results with a `two-temperature' approximation to compute the absorption \citep[as used in][]{tremblin2013}. Our models ignore self-shielding in the calculation of the radiative pressure. In the Appendix \ref{sec:ssA} we estimate the effect of self-shielding in our calculations.

\section{The model}
\label{model}

We adapted the radiation-hydrodynamics/magnetohydrodynamics code {\sc guacho} \citep{esquivel2009,2013ApJ...779..111E,villarreal2018} to simulate the interaction between the stellar wind and the photoionizing flux, with  the escaping planetary atmosphere of a hot-Jupiter around a solar type star (i.e. the HD 209458 system).

\subsection{The hydrodynamics core}
\label{sec:code}

The setup used in the code solves the ideal hydrodynamics equations with source terms due to gravity, radiation pressure, and radiative gains and losses in a Cartesian grid:
\begin{equation}
\frac{\partial \rho}{\partial t} + \nabla \cdot (\rho {\bf u})=0,
\label{eq:cont}
\end{equation}
\begin{equation}
\frac{\partial (\rho {\bf u})}{\partial t} + \nabla \cdot (\rho {\bf
  uu}+{\bf I}P)= \rho \left[{\bf g_p}+(1-\beta\chi_n){\bf g_*}\right] ,
\label{eq:mom}
\end{equation}
\begin{equation}
\begin{split}
\frac{\partial E}{\partial t} + \nabla \cdot \left[{\bf
    u}(E+P)\right]=&G_\mathrm{rad}-L_\mathrm{rad} + \\
    & \rho \left[ {\bf g_p}+ \left(1-\beta\,\chi_n\right)\,{\bf g_*}\right] \cdot
     {\bf u},
\label{eq:ener}
\end{split}
\end{equation}
where $\rho$, ${\bf u}$, $P$, and $E$ are the mass density, velocity, thermal pressure and energy density, respectively. {\bf I} is the identity matrix,  $G_\mathrm{rad}$ and $L_\mathrm{rad}$ the gains and losses due to radiation, and ${\bf g_*}$ and ${\bf g_p}$  are the gravitational acceleration due to the star and the planet, respectively. The stellar radiation pressure  is included as a decrease of  ${\bf g_*}$ by an amount proportional to $\beta$, the ratio of the radiation pressure force and the stellar gravity force, and the fraction of neutrals ($\chi_n$) within each cell. The effective gravitational acceleration that the neutral planetary material feels is then $(1-\beta\chi_n)\,\mathbf{g_*}$.

The total energy density and thermal pressure are related by an ideal gas equation of state $E = \rho \vert{\bf u}\vert^2/2 + P/(\gamma-1)$, where $\gamma=5/3$ is the ratio between specific heat capacities.

The hydrodynamics equations without the source terms are advanced with a second order Godunov method with an approximate Riemann solver (the HLLC solver, \citealt{torobook}), and a linear reconstruction of the primitive variables using the $\mathrm{minmod}$ slope limiter to ensure stability. After each temporal update the appropriate source terms are added as described below.

\subsection{Source terms}
\label{source}

Every time-step, after the hydrodynamic variables are updated with the approximate Riemann solver, the source terms (right hand side of Equations \ref{eq:cont}--\ref{eq:ener}) are computed and added to the solution in a semi-implicit scheme. In what follows, to facilitate the interpretation of the results, we discuss each term individually.

\subsubsection{Radiation Pressure}
\label{radp}

For a given transition, the ratio between the stellar radiation pressure and the gravity force can be approximated using the formula derived in \cite{lagrange}:
\begin{equation}
\begin{split}
	\beta \simeq 0.506 f \bigg( \frac{d^2}{\text{pc}^2} 		\bigg) \bigg( \frac{m}{\text{a.m.u}} \bigg)^{-1} 	\bigg( \frac{M_*}{M_{\odot}} \bigg)^{-1} \\ 			\bigg(\frac{\lambda}{2000 \textrm{\AA}} \bigg)^2
	\bigg( \frac{\Phi_{\lambda}^d}{10^{-11}\text{erg} \,			\text{cm}^{-2} \, \text{s}^{-1} \, \textrm{\AA}^{-1}} \bigg),
\end{split}
\label{eq:lagrange}
\end{equation}
where $f$ and $\lambda$ are the oscillator strength and the central wavelength of the transition, $m$ is the mass of the element in question, $M_*$ is the stellar mass,  and $\Phi_{\lambda}^d$ is the flux per unit $\lambda$ at a distance $d$.

In the XUV, the most important contribution to radiation pressure over the hydrogen atoms ($m_\mathrm{H}=1.66 \times 10^{-24}$ g) is due to the Ly$\alpha$ line ($\lambda=1214 \AA$, $f=0.4164$) and so $\Phi_{\lambda}^d$ is the intrinsic flux in Ly$\alpha$ of HD209458 ($d=47$ pc) as seen from the Earth taken from the work of \citet{bourrier2013}. Using the Ly$\alpha$ profile as a function of the radial velocity (with respect to the star) we can estimate the value of $\beta$ that enters in Eq. \ref{eq:ener} for any given cell in the computational domain.

\subsubsection{Hydrogen rate equations}
\label{sec:hrate}

In order to include the charge exchange, collisional ionization, as well as photoionization processes the modified version of the code solves, together with the hydrodynamic equations, a set of coupled rate equations (eq \ref{eq:hrateh0}-\ref{eq:hratecp}) for four hydrogen species: neutral stellar wind $H_\mathrm{h}^0$,  ionized stellar wind $H_\mathrm{h}^+$, neutral planetary wind $H_\mathrm{c}^0$,  and ionized planetary wind $H_\mathrm{c}^+$. In what follows we describe the implementation.

\paragraph{Radiative transfer}
\label{p:radT}

We consider the photoionization of H in a similar way as in \citet{schneiter}. In every cell, the temperature and with it a collisional ionization coefficient $c(T)$ and a radiative recombination coefficient $\alpha(T)$ are computed.
Photoionization rates for each of the neutral components, $\phi_\mathrm{c}$ and $\phi_\mathrm{h}$, are computed using the same ray tracing  method described in \citet{schneiter}.
The stellar ionizing flux is divided in $10^6$ photon packets which are launched in random directions from the stellar surface. The flux is attenuated along the path length as it encounters neutral material, adding to the photoionization rate at each cell. The photoionization rate  $\phi$ within each cell is then calculated by equating the ionizing photon-rate, $S_\star$, with the ionizations per unit of time within each cell ($S_\star=n_{\mathrm{H_I}}\,\phi\,\mathrm{dV}$).

\paragraph{Charge exchange}
\label{p:che}

The charge exchange of hydrogen is considered with a reaction of the form:
\begin{equation}
  \label{eq:CE}
  H_\mathrm{h}^+ + H_\mathrm{c}^0 \getsto  H_\mathrm{h}^0 + H_\mathrm{c}^+,
\end{equation}
where the subscript `h' stands for {\it hot} material (fast moving stellar wind, injected at $10^6~\mathrm{K}$), the subscript `c' denotes the {\it cold} material (slow moving planetary wind, imposed at $10^4~\mathrm{K}$).
This process allows the neutralization of fast stellar wind ions through elastic collisions with slow neutrals from the photo-evaporating planetary atmosphere.
Denoting by $\chi_\mathrm{h/c}^{0/+}$ the hot/cold, neutral/ionized fractions, the
number density of each of the species can be obtained by multiplying such fractions by the total hydrogen density $n_\mathrm{H}$. Evidently,  this implies that the ionized and neutral fractions satisfy
\begin{equation}
  \label{eq:fractions}
  \chi_\mathrm{h}^0 + \chi_\mathrm{h}^+ + \chi_\mathrm{c}^0 + \chi_\mathrm{c}^+ = 1.
\end{equation}
The hydrogen charge exchange occurs at the same rate, $\alpha_\mathrm{ce}$, in both directions of equation (\ref{eq:CE}). The reaction rate is in reality proportional to the relative velocity of neutrals and ions \citep[see][]{lindsay}. However, for the velocity range involved in our simulations the hydrogen exchange rate can be roughly approximated by a constant value $\alpha_\mathrm{ce}=4\times 10^{-8}\,\mathrm{cm^3\,s^{-1}}$ as \citet{tremblin2013}. These authors computed the evolution of the ionization state of these species by means a simple implicit method, which is valid in the absence of additional sources. In order to include other mechanisms such as ionization  and/or recombination, we implemented a chemistry module as described below.

\paragraph{Chemistry step}
\label{sec:chem}

To solve the ionization state of the species, accounting for ionization by collisions and EUV photons, radiative recombinations, and charge exchange, we include the following set of equations:
\begin{eqnarray}
  \frac{\partial \left(n_\mathrm{H}\, \chi^0_\mathrm{h}\right)}{\partial t} + \nabla\cdot\left(n_\mathrm{H} \,\chi_\mathrm{h}^0{\mathbf{u}}\right) &=
  &n_\mathrm{e}\,n_\mathrm{H}\,\left[ \chi_\mathrm{h}^+\,\alpha(T) - \chi_\mathrm{h}^0\,c(T)\right] \nonumber \\
  &-& \alpha_\mathrm{ce}\,n_\mathrm{H}^2\,\left(\chi_\mathrm{h}^0\,\chi_\mathrm{c}^+ - \chi_\mathrm{h}^+\,\chi_\mathrm{c}^0\right) \nonumber \\
  &-&n_\mathrm{H}\,\chi_\mathrm{h}^0\,\phi_\mathrm{h}~, \label{eq:hrateh0} \\
  \frac{\partial \left(n_\mathrm{H}\, \chi^+_\mathrm{h}\right)}{\partial t} + \nabla\cdot\left(n_\mathrm{H} \,\chi_\mathrm{h}^+{\mathbf{u}}\right) &=
  &-n_\mathrm{e}\,n_\mathrm{H}\,\left[ \chi_\mathrm{h}^+\,\alpha(T) - \chi_\mathrm{h}^0\,c(T)\right] \nonumber\\
  &+& \alpha_\mathrm{ce}\,n_\mathrm{H}^2\,\left(\chi_\mathrm{h}^0\,\chi_\mathrm{c}^+ - \chi_\mathrm{h}^+\,\chi_\mathrm{c}^0\right)\nonumber \\
  &+&n_\mathrm{H}\,\chi_\mathrm{h}^0\,\phi_\mathrm{h}~, \label{eq:hratehp} \\
  \frac{\partial \left(n_\mathrm{H}\, \chi^0_\mathrm{c}\right)}{\partial t} + \nabla\cdot\left(n_\mathrm{H} \,\chi_\mathrm{c}^0{\mathbf{u}}\right) &=
  &n_\mathrm{e}\,n_\mathrm{H}\,\left[ \chi_\mathrm{c}^+\,\alpha(T) - \chi_\mathrm{c}^0\,c(T)\right] \nonumber\\
  &-& \alpha_\mathrm{ce}\,n_\mathrm{H}^2\,\left(\chi_\mathrm{c}^0\,\chi_\mathrm{h}^+ - \chi_\mathrm{c}^+\,\chi_\mathrm{h}^0\right) \nonumber \\
  &-&n_\mathrm{H}\,\chi_\mathrm{c}^0\,\phi_\mathrm{c}~, \label{eq:hratec0} \\
  \frac{\partial \left(n_\mathrm{H}\, \chi^+_\mathrm{c}\right)}{\partial t} + \nabla\cdot\left(n_\mathrm{H} \,\chi_\mathrm{c}^+{\mathbf{u}}\right) &=
  &-n_\mathrm{e}\,n_\mathrm{H}\,\left[ \chi_\mathrm{c}^+\,\alpha(T) - \chi_\mathrm{c}^0\,c(T)\right] \nonumber \\
  &+& \alpha_\mathrm{ce}\,n_\mathrm{H}^2\,\left(\chi_\mathrm{c}^0\,\chi_\mathrm{h}^+ - \chi_\mathrm{c}^+\,\chi_\mathrm{h}^0\right) \nonumber \\
  &+&n_\mathrm{H}\,\chi_\mathrm{c}^0\,\phi_\mathrm{h}~, \label{eq:hratecp}
\end{eqnarray}
where $n_\mathrm{e}$ is the electron density. Assuming that the all electrons arise from the ionization of hydrogen we have the additional relation
\begin{equation}
  \label{eq:ne}
  n_\mathrm{e} = n_\mathrm{H}\,\left(\chi_\mathrm{h}^+ + \chi_\mathrm{c}^+ \right).
\end{equation}

These equations are solved with the {\sc kimya} code \citep{2018RMxAA..54..409C}, which has been incorporated as a chemistry module in {\sc guacho}. This chemistry module advances the rate equations (along with the conservation conditions in  Equations \ref{eq:fractions} and \ref{eq:ne}) with an implicit scheme that uses an iterative Newton-Raphson method.

\subsubsection{Heating and cooling}
\label{radtransfer}

The total ionization fraction (adding the contribution of both the stellar and the planetary winds) is used to calculate the radiative cooling using a parametrized curve that depends on the temperature, density and ionization state.
The cooling follows the prescription described in \citet{1995MNRAS.275..557B}.
For low temperatures ($<3\times\,10^{4}~\mathrm{K}$) the cooling due to collisional excitation and ionization of hydrogen is added explicitly.
We assume that the ionization state of oxygen follows that of hydrogen at such temperatures (where most of the oxygen is only neutral and singly ionized) and estimate the cooling due to collisional excitation of oxygen, which is added multiplied by a constant factor to account for the cooling of other elements.
At higher temperatures, where a significant amount of higher ions of oxygen should be present, we change to a coronal equilibrium cooling curve. The photon flux is used to compute a photoionization heating term \citep[see details in][]{schneiter}.

\subsection{Parameters of the simulations}
\label{parameters}
The numerical models are based on the exoplanet system HD 209458 which comprises of a solar like star and a hot Jupiter like planet.
Similarly to our previous works \citep{schneiter2007,villarreal2014,schneiter,villarreal2018} we place the source that corresponds to the star at the centre of the computational domain, which in this case corresponds to a Cartessian grid of $400\times100\times400$ ($x,y,z$) cells, with a resolution of $7.48\times 10^{9}\mathrm{cm}$ per cell. We must note that such resolution is barely sufficient to resolve the wind launching region from the planet with $\sim 8$ grid cells. With this resolution the details of the stratification of the planetary atmosphere, and its heating and photoevaporation can not be considered. Instead the planetary wind is imposed as a boundary condition and we focus our analysis in the larger (and better resolved) cometary wake and its corresponding Ly$\alpha$ absorption.

The stellar and planetary wind are reimposed at every time-step with the planet position being updated according to its orbital position (with an orbital period of $\tau_\mathrm{orb}=3.52~\mathrm{days}$).  The orbit lies in the
$xz$ plane, and it is assumed to be circular with a radius of $0.047~\mathrm{au}$, and takes place in an anti-clockwise direction. The initial position of the planet is $25\degr$ `behind' the $x$-axis to ensure that the wake is fully formed by the time it reaches the $-z$-axis, which we have taken to be the line of sight (LOS).

Both winds are launched within a fixed radius which is near the sonic point in each case, these radii ($R_\mathrm{w,p}$ and $R_\mathrm{w,*}$, respectively) are presented in Table \ref{tab:1} along with the rest of the simulation fixed parameters.

The stellar wind velocity and temperature at $R_\mathrm{w,*}$ correspond to model B in \citet{schneiter}.
The ionizing photon rate (i.e. the number of ionizing photons per unit time emitted by the star) is estimated from the stellar EUV observed luminosity in the range of $L_\mathrm{EUV} = 5.5 \times  10^{27}~\mathrm{erg~s^{-1}}$ reported by \citet{sanz-forcada2011}, and  $L_\mathrm{EUV} = 1.7 \times 10^{28}$~erg~s$^{-1}$ in
\citet{2017MNRAS.464.2396L}. The ionizing photon rate for each of the models is listed in Table \ref{tab:2}, assuming that each photon is at the Lyman limit. That is, the total EUV luminosity divided by the energy of an individual photon, which is assumed to be $13.6$ eV.
The resulting radiative stellar flux is {$F_0\simeq840$-$2\,600~\mathrm{erg\,cm^{-2}\,s^{-1}}$} at the orbital distance of HD 209458b.

\begin{table}
\caption{Stellar and planetary winds constant parameters used in simulations}
  \label{tab:1}
  \centering
  \begin{tabular}{lll}
\hline
   {\bf Stellar parameters} & Symbol & Value
    \\ \hline
   Radius$^a$               & $R_*$  &   $1.146~\mathrm{R_\odot}$    \\
   Mass$^b$                 & $M_*$  &   $1.148~\mathrm{M_\odot}$    \\
   Wind velocity        & $v_\mathrm{*}$  &  $205~\kms$\\
   Wind launch radius   & $R_\mathrm{w,*}$  &   $3.5~R_*$    \\
   Wind temperature     & $T_\mathrm{*}$  &   $1.3\times 10^6~\mathrm{K}$               \\
   Mass loss rate       & $\dot{M_*}$ &
                                                 $2.0\times10^{-14}~\mathrm{M_{\odot}\,yr^{-1}}$
    \\
   \hline
   {\bf Planetary  parameters} & Symbol  & Value
    \\ \hline
   Radius$^{c}$               & $R_\mathrm{p}$ &   $1.38~\mathrm{R_{Jup}}$    \\
   Mass$^c$                 & $M_\mathrm{p}$ &   $0.67~\mathrm{M_{Jup}}$    \\
   Wind velocity        & $v_\mathrm{p}$ &   $10~\kms$  \\
   Wind launch radius   & $R_\mathrm{w,p}$ &   $3~R_\mathrm{p}$              \\
   Wind temperature     & $T_\mathrm{p}$ &   $1 \times 10^4~\mathrm{K}$    \\
    \hline
    \multicolumn{3}{l}{$^a$ \citet{brown2001}}\\
    \multicolumn{3}{l}{$^b$ \citet{Southworth2010}}\\
    \multicolumn{3}{l}{$^a$ \citet{mazeh2000}}\\
  \end{tabular}

\end{table}

\begin{table}
 \caption{Parameters varied in the simulations}
 \label{tab:2}
 \centering
  \begin{tabular}{lccc}
\hline
{\bf Model} & $\dot{M}_\mathrm{p}$& Planet neutral                   & Ionizing photon  \\
            &                           & fraction                         & rate             \\
            & $(\times 10^{10}~\mathrm{g\,s^{-1}})$& $\chi_\mathrm{c}^0$&  $(\times 10^{38}~\mathrm{s^{-1}})$  \\ \hline
M1        & 1.5 & 0.5  &  2.4  \\
M2        & 1.5 & 0.2  &  2.4  \\
M3  & 1.5 & 0.5  &  7.4 \\
M4  & 1.5 & 0.2  &  7.4 \\
M5  & 3.0 & 0.5  &  7.4 \\
M6  & 3.0 & 0.2  &  7.4 \\
\hline
  \end{tabular}
\end{table}

For the planetary wind we set the values of temperature and velocity using the results of the atmospheric escape model of \citet{murray-clay2009}, which is also in agreement with the recent model of \citet{Salz2016}. This model predicts a ionization fraction of $0.2$ at $3~R_\mathrm{p}$, which corresponds to the distance at which we impose the planetary wind. This fraction is set for models M2, M4 and M6 (see Table \ref{tab:2}). A neutral fraction of $0.5$ is also explored in models M1, M3 and M5, corresponding to the results obtained by \citet{koskinen2013}.

The adopted mass loss rate range is consistent with the values used in our previous works, and  also with those obtained in \citet{tripathi2015} and \citet{2017MNRAS.464.2396L}.
The velocity, mass loss-rate, and temperature are taken to be constant, the resulting density profile is obtained from $\rho=\dot{M}_\mathrm{s,p}/(4\pi R_\mathrm{s,p}^2 v_\mathrm{s,p})$ inside the wind imposing region for the star (s) or the planet (p).

\section{Results}\label{results}

\begin{figure*}
  \centering
  \includegraphics[width=0.33\textwidth]{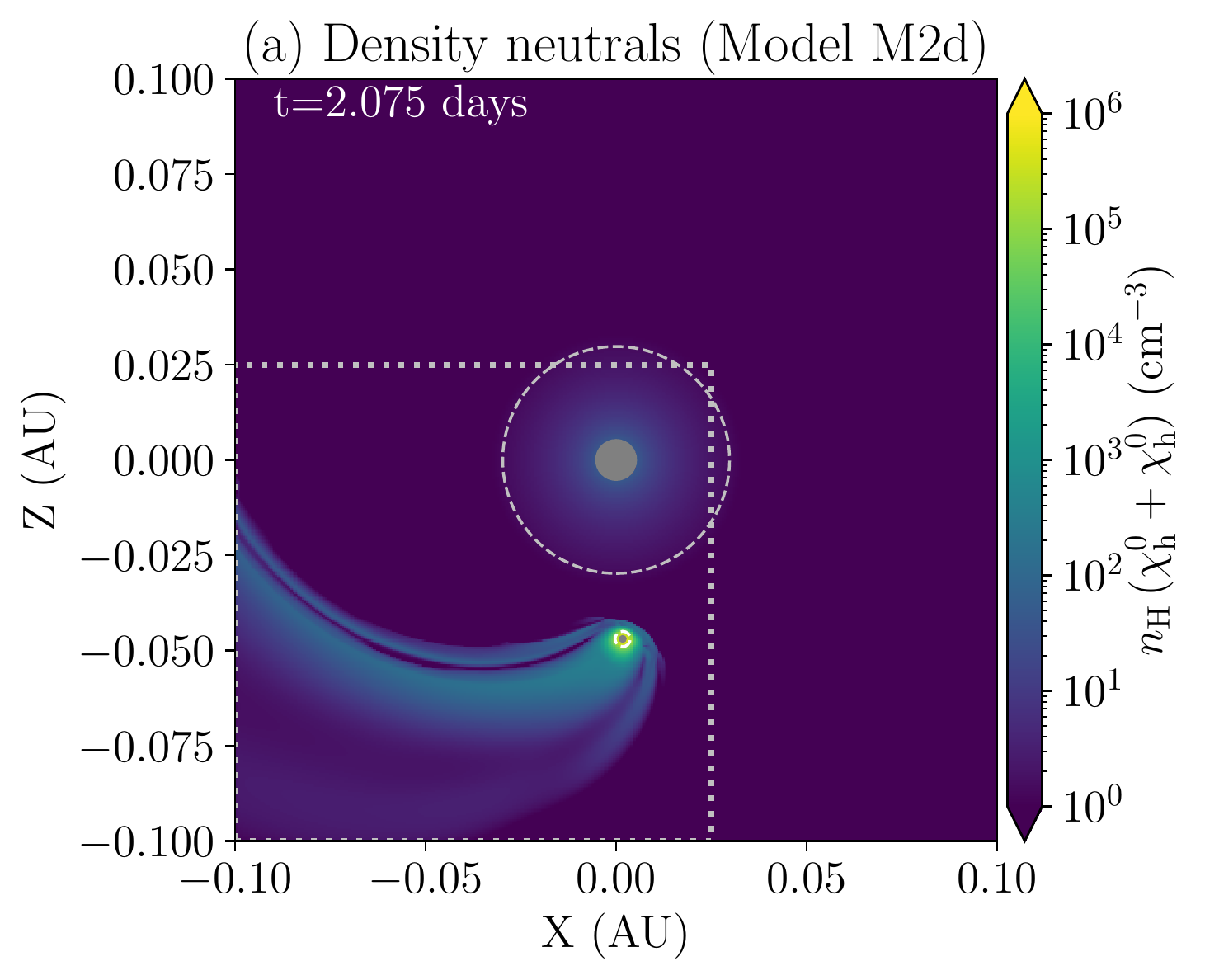}
  \includegraphics[width=0.33\textwidth]{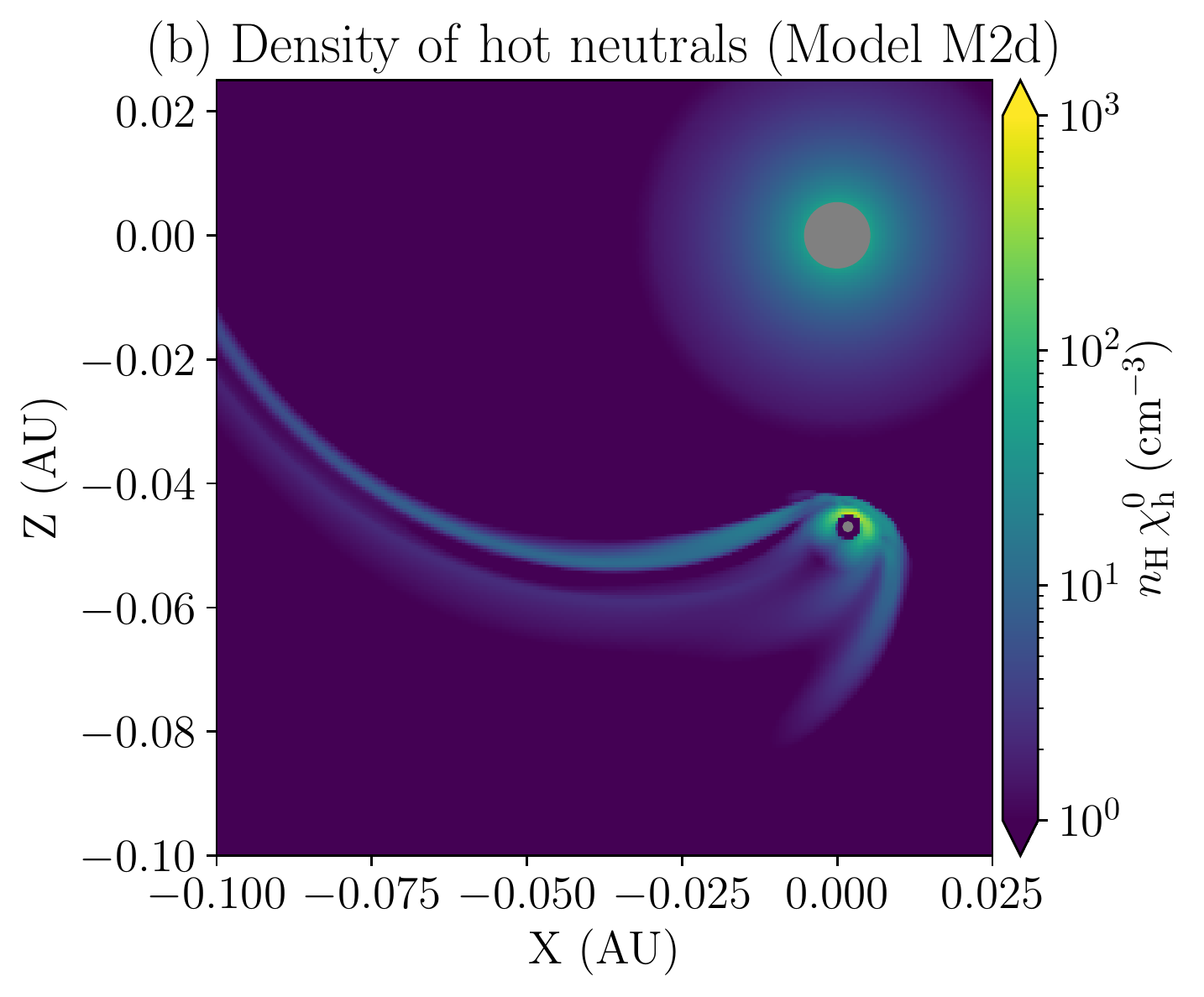}
  \includegraphics[width=0.33\textwidth]{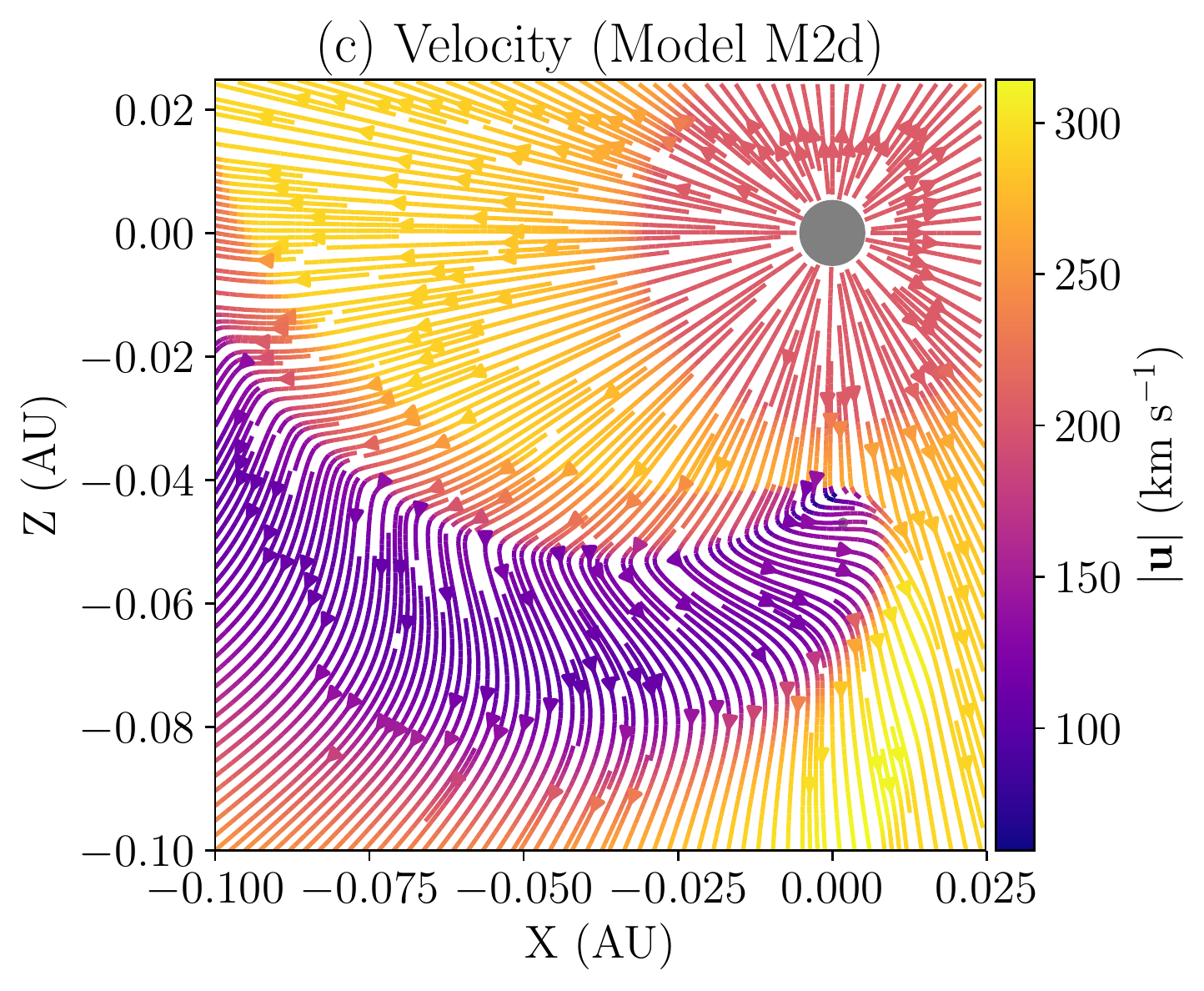}
  \includegraphics[width=0.33\textwidth]{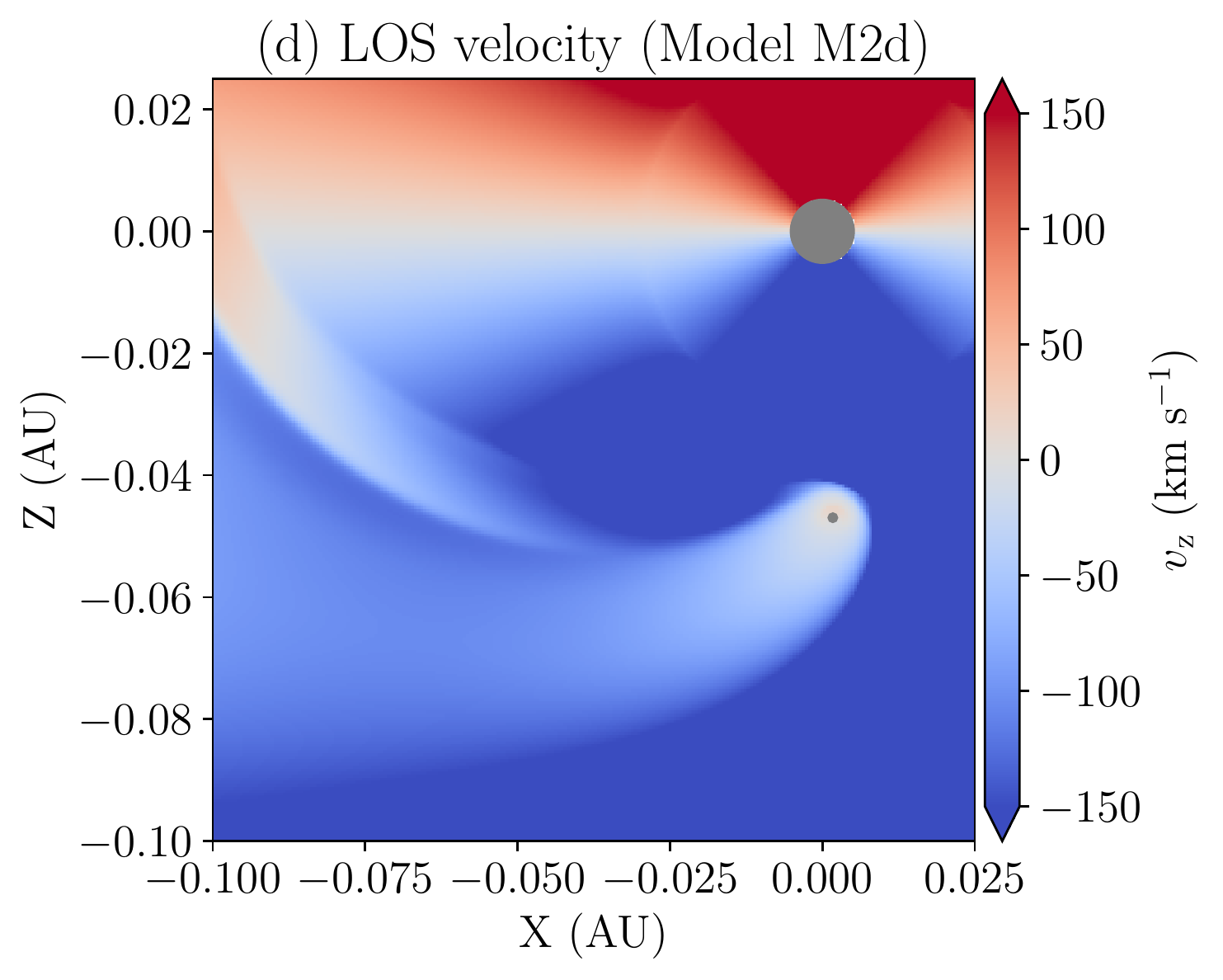}
  \includegraphics[width=0.33\textwidth]{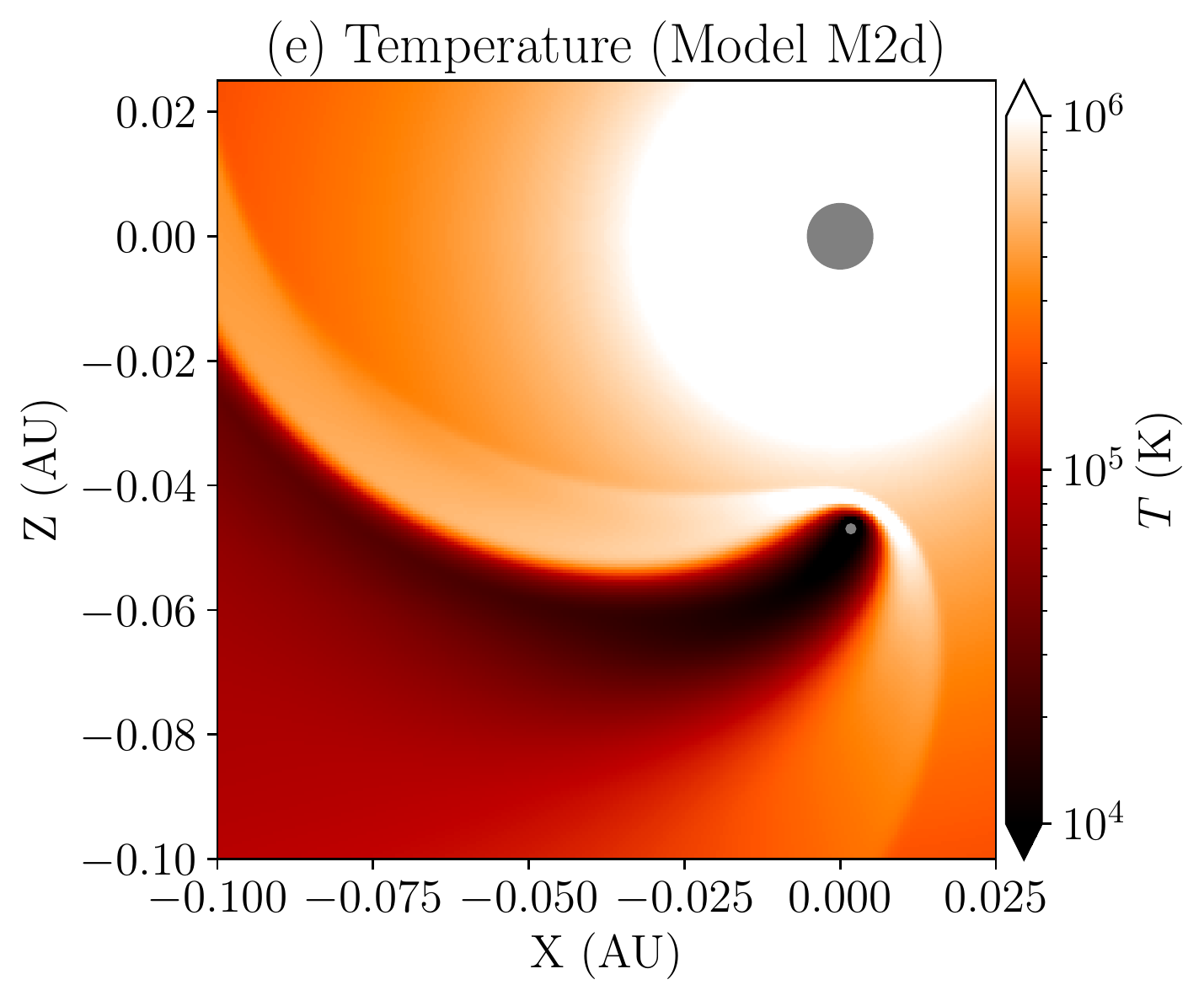}
  \caption{Example of the flow configuration in the orbital plane for one of the Models (M2d). In panel (a) we show a cut of the entire domain with the stratification of neutral density after an integration time of $t=2.075~\mathrm{days}$, at which the planet is in transit if observed from $-z$. The regions that correspond to the star and planet are shaded in grey, and with a circle in dotted lines we show the regions from which we impose the winds. In the following panels we show the inner region marked by a rectangle in (a). In (b) the density of neutrals from the stellar wind that were neutralized by charge exchange. The velocity field in the orbital plane is shown as stream-lines in panel (c); the magnitude of $v_\mathrm{z}$ (approximately the LOS velocity) is in panel (d); and the temperature in (e). The values and units used in the plots are shown in the color-bar the right of each one.}
  \label{fig:2dmaps}
\end{figure*}

In order to study how the processes of charge exchange and radiation pressure affects the blue part of the Lyman $\alpha$ line during the transit of HD 209458b, we implement these two mechanisms in the simulations. Since they depend on the ionization fraction and the ionizing flux from the host star, we vary  these two parameters as shown in Table \ref{tab:2}.
To separate the effects that charge exchange and the stellar radiation pressure have on the neutral planetary material we run, for every model listed in Table \ref{tab:2}, a total of four simulations or sub-models, that we denote by an additional letter: `a' with neither charge exchange nor radiation pressure enabled, `b' with charge exchange but no radiation pressure, `c' with radiation pressure but no charge-exchange, and `d' with both charge exchange and radiation pressure. To turn off the charge-exchange in models `a' and `c' we set the rate coefficient $\alpha_\mathrm{ce}$ in Equations (\ref{eq:hrateh0})--(\ref{eq:hratecp}) to zero. Similarly, models without radiation pressure are obtained with $\beta=0$ in Equations (\ref{eq:mom}) and (\ref{eq:ener}).

The results of the simulations include the density of neutrals (from both the stellar and planetary winds), the (3D) velocity fields, and the temperature distribution (obtained from the thermal pressure and density). In Figure \ref{fig:2dmaps} we show an example of the flow configuration in the orbital plane at the time of maximum absorption (at $t=2.075~\mathrm{days}$, with the planet in full transit if observed from $-z$). In the figure we show only one model (M2d), but all of them are qualitatively similar with slight differences. For instance, a higher planetary mass loss rate would result in a broader wake, and a higher ionization fraction would change the amount of absorbing material. However, these details are not very evident from the flow configuration itself, but can be seen in the Ly$\alpha$ absorption profile.

These profiles are computed as a function of velocity at the orbital position \citep[see][]{schneiter}, considering an orbit with an inclination of ${i=86.59\degr}$ (the orientation of HD~209548b).

\begin{figure*}
  \centering
  \includegraphics[width=\textwidth]{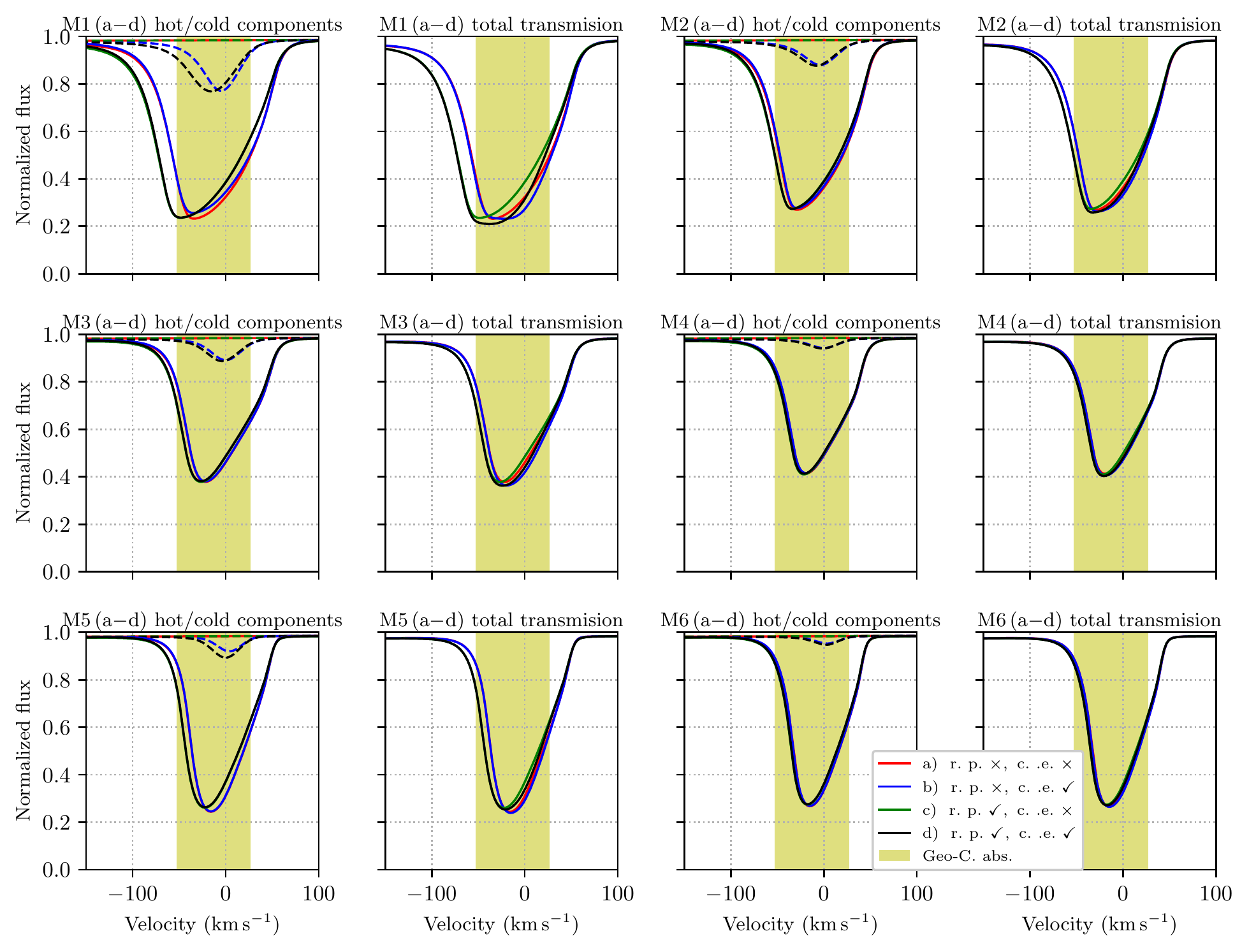}
  \caption{Ly$\alpha$ transmission in transit, as a function of LOS velocity. The columns labelled `hot/cold components' (leftmost and second to last) are obtained considering only the opacity of the cold neutrals (solid lines), and only the opacity of the hot neutrals (dashed lines). The columns labelled `total transmission' are obtained considering the opacity of both hot and cold components. The model used for each plot is indicated in the title, and the sub-models are coded by the line colour as indicated in the legend at the bottom/right of the figure.}
  \label{fig:abs_vs_vel}
\end{figure*}

In Figure \ref{fig:abs_vs_vel} we show the Ly$\alpha$ transmission as a function of the velocity at the moment of highest absorption (with the planet in transit). In the first and third columns from the left in the figure, we show the separate contribution to the observed absorption produced by neutrals in the stellar wind (dashed lines), and the neutrals from the planetary wind (solid lines), In the second and fourth columns we show the combined transmission after absorption of both neutral components. The different colours in the lines correspond to the different submodels (a-d), as show in the legend of the figure.
The shaded area in the plots corresponds to a region that is excluded in the analysis of the observations because it is affected by geo-coronal emission \citep{vidal2003}.

The absorption for each component is obtained by calculating the line profile along the LOS, in the directions that intercept the stellar disk. To obtain the line profiles we use the temperature and density of neutrals in each cell. In our simulations we treat the two neutral species as components in a single fluid. Thus they have the same temperature, although the ionization state is computed separately out of equilibrium.
However, we can set a different temperature for each population in post-processing to compute the Ly$\alpha$ transmission profile as in \citet{tremblin2013}. These authors consider an alternative treatment in which they impose a different temperature for the stellar wind ($10^6~\mathrm{K}$) and the planetary wind ($7\,000~\mathrm{K}$), but instead of including an explicit treatment of heating and cooling processes they use an effective adiabatic index $\gamma = 1.01$. With such equation of state the flow tends to preserve its original temperature, and it is only different at the interaction region.
In our models we use an adiabatic index of $\gamma = 5/3$, but we include heating and cooling explicitly. Our treatment of heating and cooling is adequate for the planetary wind, which is the main contribution to the absorption. We do not, however, include thermal conduction which is responsible to maintain an almost isothermal stellar wind. As a result the stellar wind temperature falls with distance from the star (see Fig \ref{fig:2dmaps}e) more than in reality. This would result in an underestimation of the thermal pressure of the stellar wind at the planet position, however at such distance the ram pressure is dominant.

To compare with the observatioal Ly$\alpha$ absorption of HD~209458b in \citet{vidal2003} we present synthetic observed profiles in Figure \ref{fig:flux_vs_VM} (and \ref{fig:flux_vs_VM-TT} in the appendix \ref{sec:ttA}).
These synthetic profiles are obtained taking the intrinsic (un-absorbed) profile constructed by \citet{bourrier2013}, then we include the absorption produced by neutral hydrogen along the LOS from the planet evaporating atmosphere, and that of the ISM\footnote{The ISM absorption is computed considering absorption from hydrogen and deuterium as in \citet{wood2005} with $N_\mathrm{HI}=10^{18.37}~\mathrm{cm^{-2}}$, $N_\mathrm{DI}=1.5\times 10^{-5}~N_\mathrm{HI}$. And two Voigt profiles with doppler broadening parameters $b_\mathrm{HI}=12.3~\kms$ and $b_\mathrm{DI}=b_\mathrm{HI}/\sqrt{2}$, centered at $-6.6~\kms$ for \ion{H}{1}, and $-87.6~\kms$ for \ion{D}{1}.} In order to make a fair comparison with the observations we convolve the results with the STIS G140M instrumental response with the resolution used by \citet{vidal2003}.

\begin{figure*}
  \centering
  \includegraphics[width=\textwidth]{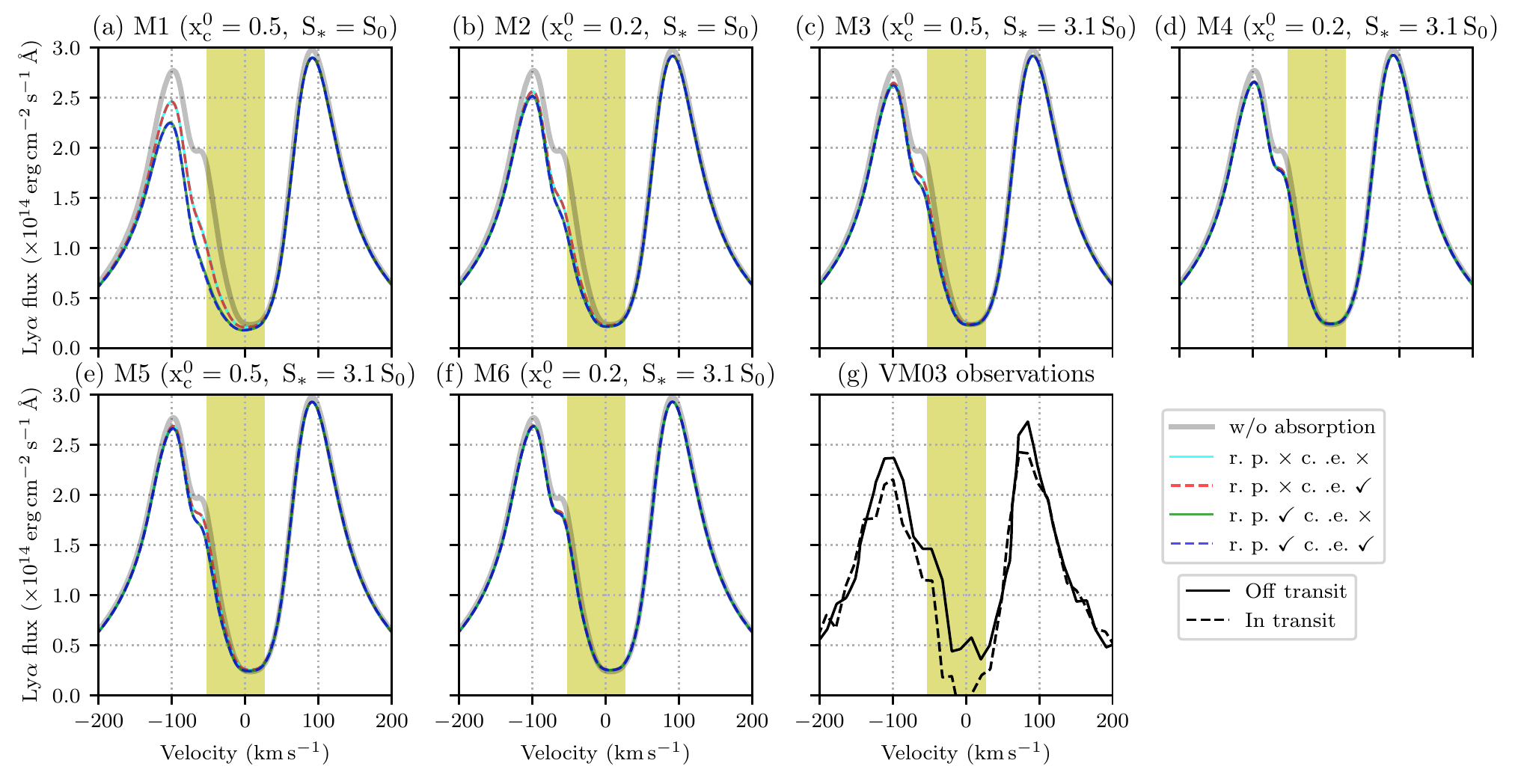}
  \caption{Synthetic observations obtained from the absorption computed in our models and the  Ly$\alpha$ observations reproduced form \citet[][]{vidal2003}. In the models (panels a--f) we show in grey the observed emission un-absorbed by the exoplanet atmosphere (only ISM absorption), and in different colors the submodels, as indicated in the label. Panel (g) shows the observations in and off transit, dashed and solid lines, respectively.}
  \label{fig:flux_vs_VM}
\end{figure*}

\section{Discussion}\label{discussion}

We can see in Figure \ref{fig:abs_vs_vel} how the Ly$\alpha$ absorption is affected by the treatment of the radiation pressure, and the inclusion of charge exchange between the two winds. In what follows we discuss these effects, and we compare with the observations of the HD~209458 system.
The first thing to notice is that both processes have a rather marginal effect in our simulations. This seems in contradiction with previous studies, in particular those relying in particle/kinetic simulations \citep[e.g.][]{bourrier2013, kislyakova2014}. The reason is that in their case the interaction is assumed to be collisionless, while the system is not entirely so \citep[see][]{koskinen2013,Salz2016}.  The result is that the coupling between the stellar wind and the escaping atmosphere is underestimated. Our hydrodynamical simulations on the other hand tend to overestimate such coupling by treating them as a fluid. With that in mind let us analyze how charge exchange and radiation pressure affect the hydrodynamical description of the problem.

\subsection{Effects of charge exchange and temperature dependence}\label{sec:disc_ce}

Charge exchange occurs mainly at the bow-shock region where the stellar and planetary material get mixed. However, precisely because it occurs at the interaction region, in a single fluid hydrodynamical simulation \citep[as used here and in][]{tremblin2013}, the radial velocity of stellar hot neutrals in the region is greatly decreased as the result of the hydrodynamic shock.
This can be seen in Figure \ref{fig:2dmaps}{\it c} where we can see that at the interaction region there is a stagnation region of the flow, which moves as the planet travels in its orbit.
A similar behaviour can be seen in Figure \ref{fig:2dmaps}{\it d}, where we show the $v_z$ velocity component which coincides with the LOS velocity if observed from $-z$.

In Figure \ref{fig:2dmaps}{\it b} we show the distribution of hot neutrals in the orbital plane, in a region close to the planet. The snapshot was taken at the time the planet is in transit. From the figure we can see that most of the neutralized stellar wind material is located in the bow shock region, with LOS velocities of only  $\sim50~\kms$ (see Figure\ref{fig:2dmaps}{\it d}).
This can also be seen in Figure \ref{fig:abs_vs_vel} (first and third columns) where most of the absorption due to hot neutrals is at low velocities. Some higher velocity hot neutrals are present in the region behind the planet, however at a lower density. There are also some hot neutrals in the wake on the side that faces the star, but have low density as well.

A possibility that enables a larger fraction of hot neutrals to keep a high velocity in a single fluid simulation, is to consider that instabilities (such as Kelvin-Helmholtz, K-H) feed a mixing layer sliding on the sides of the bow-shock \citep{tremblin2013}.
However, in our models, the direction along the sheath of the bow-shock, as the planet transits the star, has a large angle with the line of sight, and the resolution used is insufficient to reproduce the K-H instabilities \citep[which was feasible in the 2D models of][]{tremblin2013}.
We must note however, that in our models the centre of the line is shifted to the blue because of the orbital speed of the planet, along with a whiplash effect at the neutral wake.

The resulting absorption by hot neutrals is shown in dashed lines in Figure \ref{fig:abs_vs_vel} (first and third column from the left). From the figure we can readily see that the absorption of hot neutrals is overwhelmed by that of cold neutrals (planetary material, solid lines). We can also see that the center of the absorption produced by hot neutrals is close to zero velocity, with only a small shift to the blue. As mentioned before, this is due to the shock that is formed by the two winds, bringing the stellar wind to a stop at the stagnation point. It can be noted as well that the relative contribution of hot neutrals decreases as we increase the ionizing flux from the star. This can be attributed to the fact that the hot neutrals are formed in the region more exposed to the ionizing radiation of the star. At the same time, there are only small differences in the total transmission profile (Figure \ref{fig:abs_vs_vel}, second and fourth columns from the left) between the models with charge exchange (red and green lines) with those without charge exchange (in blue, and black). Moreover, the only noticeable differences are at low velocities, in the region that is unavailable from observations.

The bow-shock has the additional effect of raising the temperature of the planetary material (see Figure \ref{fig:2dmaps}{\it e}), which translates into a broad Ly$\alpha$ absorption profile, with significant wings at $\sim -100\,\kms$. Certainly, the post-shock temperature is high enough to produce collisional ionization, leaving only a small fraction of neutrals exposed to the stellar radiation, which are further ionized by EUV photons. But, in spite of this, our models with a low ionization rate (M1 and M2) show that the surviving neutral planetary material (solid lines) still dominates the absorption at such velocities (see Figure \ref{fig:abs_vs_vel}). We also see that the absorption profile is narrower in the models with high ionization rate, which is due to the fact that a larger ionizing flux is able to penetrate the shocked planetary wind region where the temperature is higher, leaving only cooler material to absorb the Ly$\alpha$ emission from the star.

Since the width of the absorption profiles changes significantly between the models, and depends on the absorbing material temperature, we explore the calculation of the absorption with a similar approximation as done in \citet{tremblin2013} in the Appendix \ref{sec:ttA}. That is, to consider two temperatures ($10^6$ and $10^4~\mathrm{K}$, for the stellar and planetary wind, respectively) to compute the absorption profiles.
The absorption show the same general trends, but produce a lower absorption at $-100~\kms$ than those obtained with the single temperature in our simulations.
These differences, unfortunately, lie again in the region that can not be probed by observations because the geo-coronal emission. What our models show is that the absorption at high blue shifted velocities is dominated in general by planetary material with a broad thermal width that is the direct result from the hydrodynamic shock.
In order to properly address this issue in a hydrodynamic description a two-fluid approach must be taken.

\subsection{Radiation Pressure}\label{sec:disc_rp}

As mentioned earlier, the radiation pressure has been implemented using Equation \ref{eq:lagrange}, with the Ly$\alpha$ flux calculated by \citet{bourrier2013}. An important consideration is that, given the nature of the hydrodynamic approximation, we compute the neutral fraction within a cell. In consequence, the resulting force (see Eqs. \ref{eq:mom} and \ref{eq:ener}) is applied to all material within a cell, instead of to individual particles. As a result the force acts mostly in the planetary wind, and the wind interaction region, where the density of neutrals is the highest.
Another difference with the implementation of \citet{bourrier2013} is that we simplified our calculations by neglecting self-shielding within the $\beta$ profiles. A discussion about the importance of the self-shielding in our models is given in Appendix \ref{sec:ssA}.

Comparing the results with radiation pressure (green and black lines in Figure \ref{fig:abs_vs_vel}) with those without radiation pressure (red and blue lines) we can see that the resulting profile shapes are very similar. However, the inclusion of radiation pressure results in a shift to negative velocities. This behaviour is seen in all models, but becomes less pronounced as the stellar neutral density decreases, either by injecting the planetary wind with a higher ionization fraction, or by increasing the photo-ionizing flux.

The shift to blue velocities is insufficient to explain the absorption seen in Ly$\alpha$ at velocities as high as $-100~\kms$ if a profile with a narrow thermal width (e.g. with $T\sim10^4~\mathrm{K}$) is assumed.
This is in contrast with the work of \citet{bourrier2013}, who found that the radiation pressure was the principal process responsible for the absorption in the blue part of the line. These authors, measured a photon flux that is similar to that used in models M3--M6, even though in their models is reduced due to self-shielding, which they included in the calculation.
By not considering self-shielding in the $\beta$ profile in our simulations we are effectively  overestimating the force due to radiation pressure, which would make even more difficult to account for the high velocity absorption in our models by radiation pressure alone.
We show in Appendix \ref{sec:ssA} that including the self-shielding in our models, would likely make an appreciable difference, but only in models where radiation pressure is important to begin with (i.e. only M1 and lower, ionizing fluxes). This is because the same photon flux that provides the radiation pressure ionizes the planetary wake before is accelerated to the values obtained in kinetic simulations.

For a given planetary mass-loss rate and ionizing flux, models with a higher neutral fraction (M1, M3 and M5) show a higher shift to the blue from the radiation pressure than the models with a lower neutral fraction in the planetary wind (M2, M4 and M6), as well as a higher total absorption. Naturally, for a given stellar flux, models with a larger planetary mass loss-rate produce a larger total absorption.

Another interesting feature seen in our models is a faint (but perhaps measurable) absorption that persists for up to $~20~\mathrm{hrs}$ after the transit, due to the neutral wake left by the planet \cite[see also][]{schneiter}. This total absorption decays with time (depending of the models) from a range of $\sim 8$--$2$ percent (for models M1 and M6, respectively) before it becomes undetectable. However, we can not pursue a further analysis because in some of the models the tail of the wake leaves the computational domain before it is fully ionized, thus it would require to run the simulations in a larger computational box.

\section{Summary and conclusions}
\label{sec:conclusions}

We present hydrodynamical simulations of the transit of HD\,209458b, that include photoionization, collisional ionization and radiative recombinations. Our simulations also include an approximation for the radiation pressure in the Ly$\alpha$ line, as well as charge exchange between stellar and planetary material. The aim of the present work is to assess the relative contribution of the radiation pressure, and charge exchange to the observed line profiles, in particular for the $10\%$ absorption in the blue part of the spectrum, at velocities near $-100~\kms$ in hydrodynamical simulations. We use a single set of stellar wind parameters based on observations and previous studies fit for HD\,209458 (see Table \ref{tab:1}). In order to draw general conclusions for other systems a more extensive parameter study is required. Nonetheless, the present models remark the important issue of the inherent differences between hydrodynamic and kinetic models of exoplanet escaping atmospheres and their interaction with the host stellar wind.

In previous models \citep{schneiter2007,villarreal2014,schneiter,villarreal2018} radiation pressure have been included as a uniform reduction of the stellar gravity by a factor of $0.3$ \citep{vidal2003}. In the present paper we include a somewhat more refined treatment following the work of \citet{bourrier2013} which calculate the reduction of gravity as a function of radial velocity. We compare these models with the case in which the gravity of the star is not reduced at all.
We find that the radiation pressure does not provide a significant acceleration of material within the hydrodynamical simulations. The model that produces the highest blue-shift was the one with the smaller stellar ionizing flux, and higher neutral fraction in the planetary wind (M1). The acceleration by radiation pressure in that model resulted in a blue-shift of the absorption of only $\sim20~\kms$, and produced a $\sim 20\%$ absorption at $-100~\kms$.

Charge exchange has been included in kinetic \citep[e.g.][]{bourrier2013}, and hydrodynamic \citep{tremblin2013} simulations. The kinetic approach has some limitations, in particular to reproduce the shock-structure that results from the interaction of the stellar and planetary winds.
The hydrodynamic perspective is not without important challenges, as the single fluid approach forbids a proper estimation of the temperature of the two interacting (stellar and planetary) winds needed to calculate the Ly$\alpha$ absorption.
The shock in our hydrodynamical simulations heats the planetary wind has and produces a broad absorption profile, which could help explain the observations in the blue side of the line if the shift due to radiation pressure is added.

When computing the absorption profile with a two-temperature appropriation as in \citet{tremblin2013} we see that the hot neutrals provide absorption at high velocities. But, this absorption is not enough to reproduce the observations unless a significant blue-shift due to radiation pressure is included.

We must note also that when we take into account the STIS instrumental response to produce synthetic observations in Figure \ref{fig:flux_vs_VM} that could be compared with Ly$\alpha$ transit observations \citep{vidal2003} the absorption is spread in velocity and most of the detailed differences between the models are lost.

While neither radiation pressure alone, nor charge exchange by itself can reproduce the absorption at high negative speeds, we find that the combinations of both might. It is important to emphasize that this conclusion applies to a hydrodynamical model, in which a shock is formed by the two-wind interaction. Such a shock is responsible for heating (and ionize further) the neutral material coming from the evaporating atmosphere of the planet, and in fact, such shock heating is responsible for a broader absorption which produces also absorption at high blue-shifts, at least at the same level as the other two mechanisms (radiation pressure and charge exchange).

Both issues, the correct temperature determination, and the details of the interaction region would benefit from a proper two-temperature treatment, which we plan to pursue in a future work.

\section*{Acknowledgments}

AE, acknowledges support from DGAPA-PAPIIT (UNAM) grants IN109715, IG-RG 100516, and CONACYT grant
167611. MS acknowledges the financial support from the PICT project 2566/OC-AR and the
fellowship grant BECA EXTERNA AGOSTO 2013 given by CONICET. CVD acknowledges Irish Research Council postdoctoral fellowship, project 208066, award 15350.

\appendix

\section{Two-temperature approximation}
\label{sec:ttA}

The absorption profiles presented in Figures \ref{fig:abs_vs_vel} and \ref{fig:flux_vs_VM} were obtained with the density of each of hot neutrals (from the stellar wind), and cold neutrals (from the planetary wind). However, the distinction between hot and cold is somewhat arbitrary. These two components are imposed in the wind sources at different temperatures and ionization state, but since they are treated only as different species within the same fluid, as they mix they reach a common temperature.
An alternative description within a single-fluid treatment was used in \citet{tremblin2013}, with an almost isothermal equation of state (a $\gamma=1.01$). In such approximation the winds tend to preserve their original temperature, and only within a narrower mixing region the temperature is intermediate. Then, to compute the absorption only two temperatures are used to obtain the line profiles, $10^6~\mathrm{K}$ for the material from the stellar wind, and $7\,000~\mathrm{K}$ for the cold planetary wind.

\begin{figure*}
  \centering
  \includegraphics[width=\textwidth]{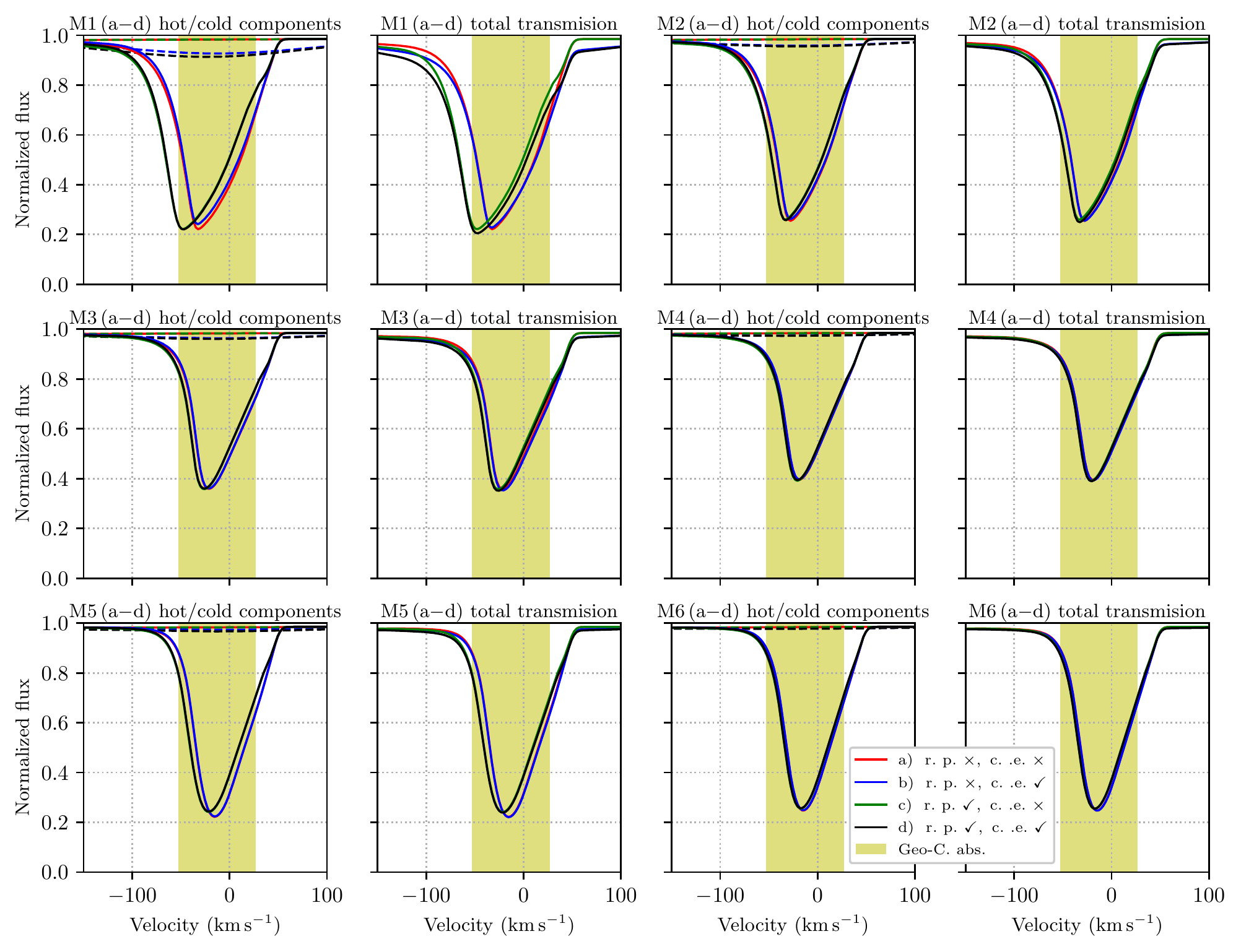}
  \caption{Same as Figure \ref{fig:abs_vs_vel}, but with the absorption computed with the two-temperature approximation, see text for discussion.
  \label{fig:abs_vs_vel-TT}}
\end{figure*}

\begin{figure*}
  \centering
  \includegraphics[width=\textwidth]{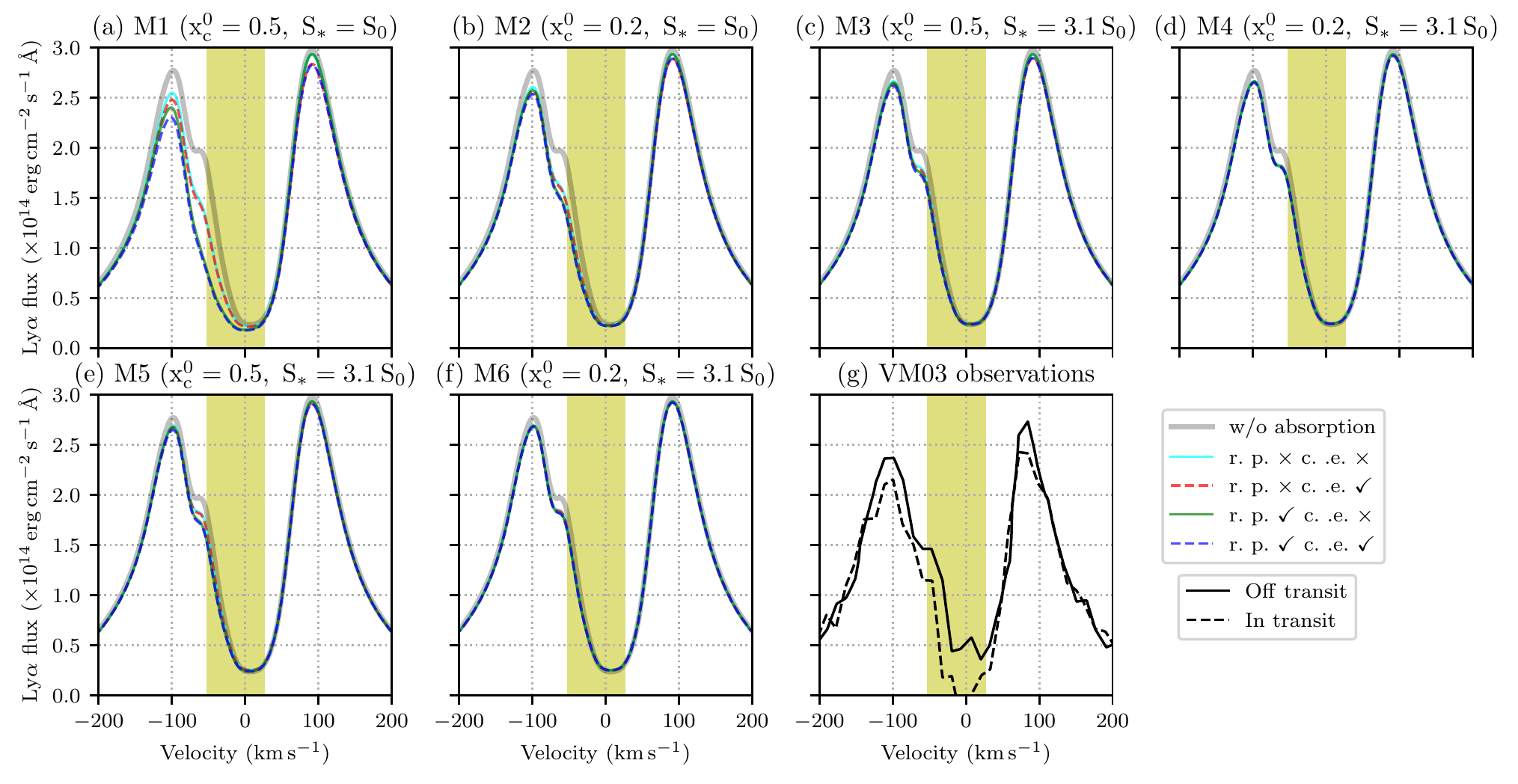}
  \caption{Same as Figure \ref{fig:flux_vs_VM}, but with the absorption computed with the two-temperature approximation.}
  \label{fig:flux_vs_VM-TT}
\end{figure*}

In this appendix we explore the effect of using two distinct temperatures for the material from each of the two neutral species (stellar and planetary). We re-compute the absorption using $10^6~\mathrm{K}$ for the line profiles with the stellar wind material, and $10^4~\mathrm{K}$ for the planetary wind. The results for this two-temperature appropriation are shown in Figures \ref{fig:abs_vs_vel-TT} and \ref{fig:flux_vs_VM-TT}.


The transmission profiles as a function of velocity (Figure \ref{fig:abs_vs_vel-TT}) are similar, but narrower than those with the temperature obtained within our simulations (Figure \ref{fig:abs_vs_vel}). We can see that the absorption from hot neutrals is negligible when charge exchange is disabled. In the sub-models where it is enabled (`b' and `d', blue and black lines) a much broader profile arises from the hot neutrals, but at a very small depth in absorption. The cold neutrals on the other hand are narrower and more peaked with a single $10^4~\mathrm{K}$ temperature, and only model M1 shows an appreciable absorption at high blue-shifted velocities.

With the two-temperature transmission profiles we also construct synthetic observations, analogous to Figure \ref{fig:flux_vs_VM} obtained with the original temperature profile. These synthetic profiles are shown in Figure \ref{fig:flux_vs_VM-TT}. We see that the differences that were noticeable in the transmission profiles, after the convolution with the STIS instrumental response become unnoticeable.

\section{Self-shielding}
\label{sec:ssA}

The radiation pressure included in our numerical models assume that the stellar flux is the same for all hydrogen particles at a given radial velocity. A more accurate profile should include the absorption of the stellar flux by the material at comparable velocity between a fluid parcel and the star. This implies the computation of a radial velocity dependent column density for each cell.
In this section we show that the shielding described by \citet{bourrier2013} would impact in our models.
The ratio of the forces due to radiation pressure and gravity acting on a parcel of gas at a given radial velocity $v_\mathrm{r}$ with respect to the star can be approximated as \citep{bourrier2013}
\begin{equation}
  \beta\left(v_\mathrm{r}, \ell \right)= \beta\left(v_\mathrm{r}, 0 \right)\,\mathrm{e}^{-\tau},
  \label{eq:beta_betau}
\end{equation}
with an optical depth
\begin{equation}
  \tau(v_\mathrm{r, \ell}) = \frac{\sigma_\mathrm{v,0} \lambda_0}{\Delta v}\int_0^{\ell} n_\mathrm{H}\left(v_\mathrm{r},\ell' \right)\,d\ell'.
  \label{eq:tau}
\end{equation}
Where the optical depth is computed integrating the column density from the cell ($\ell=0$) to the star ($\ell'=\ell$) binned in velocity intervals of with $\Delta v$; $\lambda_0$ is the wavelength of the Ly$\alpha$ line ($1215.67~\AA$), and $\sigma_\mathrm{v,0}=0.01102~\mathrm{cm\,s^{-1}}$ the effective cross section of the line  \citep{osterbrock},including the normalisation factor assuming that the line profile is a delta function centred at a velocity $v_\mathrm{r}\pm\Delta v/2$.

\begin{figure*}
  \centering
  \includegraphics[width=0.45\textwidth]{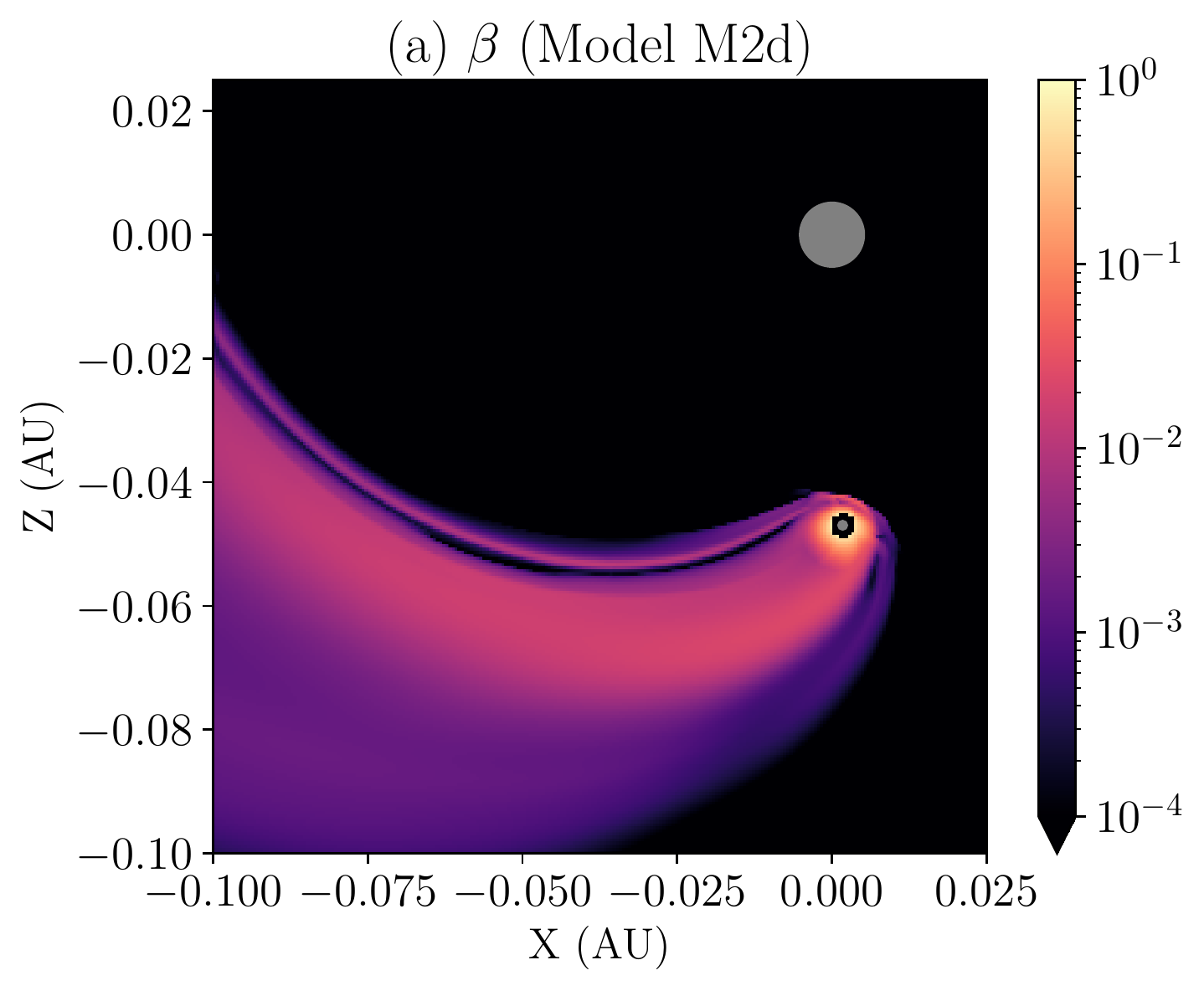}
  \includegraphics[width=0.45\textwidth]{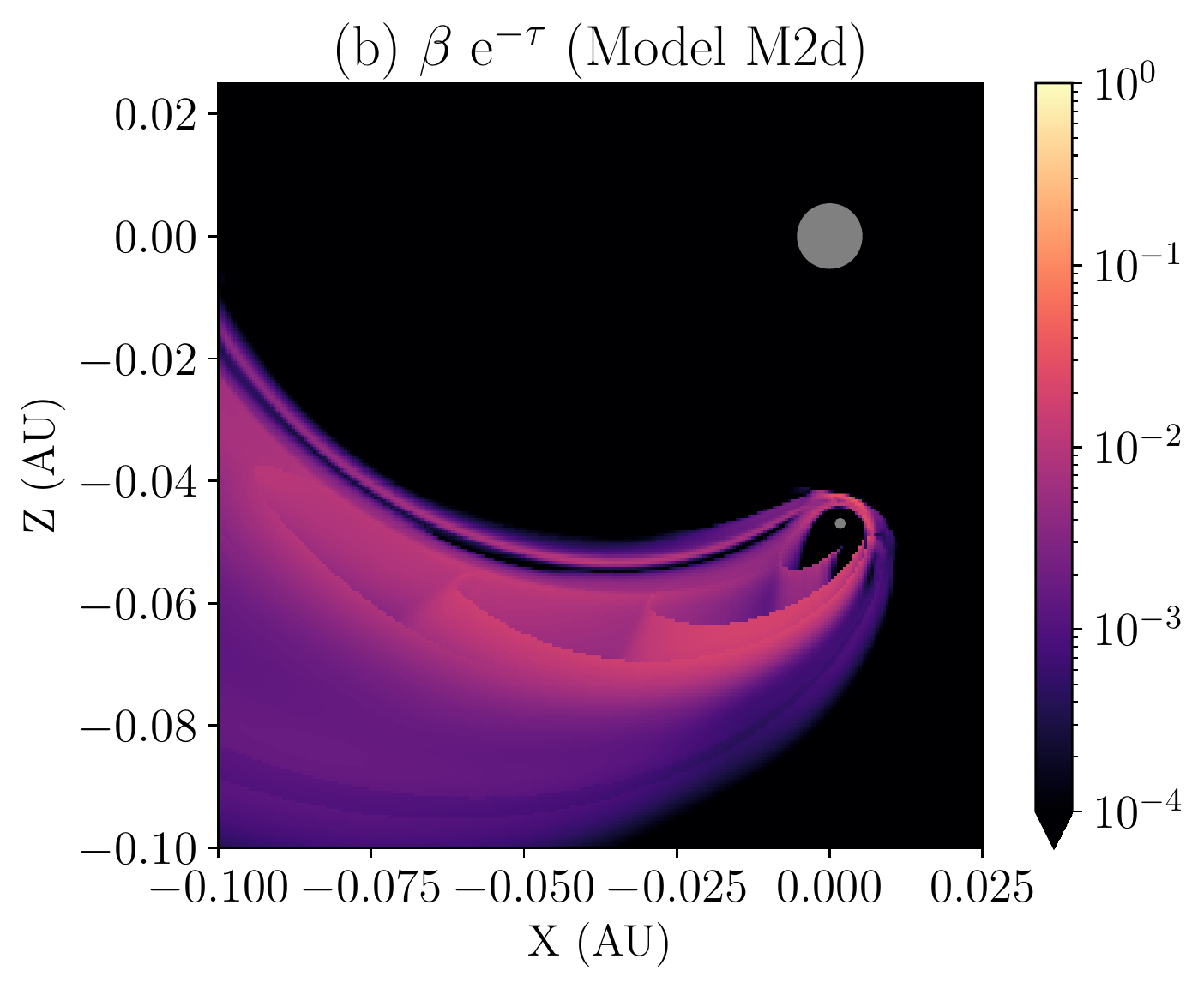}
  \caption{{Value of $\beta$ used in one of the models (M2d), in the orbital plane for the same zoom-in region shown in the Figure \ref{fig:2dmaps}. In panel (a) without self-shielding, and in panel (b) including self-shielding (only in post-processing).}}
  \label{fig:betau}
\end{figure*}

In Figure \ref{fig:betau} we show the $\beta$ value for the same model shown in Figure \ref{fig:2dmaps} (M2d) in the left panel ({\it a}). In the right panel ({\it b}) we show the resulting $\beta$ after considering the self-shielding, which was computed with a $\Delta v = 15~\kms$.

We can see from the Figure that the material closest to the planet is well shielded from the stellar radiation, and the resulting pressure. This means that to properly include the radiation pressure one would need to take into account the radiation transfer described above, not as post-processing, but within the simulation. This is possible with a grid code, but rather computationally expensive, and would require a different treatment of the radiation. In any case, by neglecting the self-shielding we are over estimating the effect of radiation pressure in the hydrodynamical simulations. A more precise account for the radiation pressure in our simulations would make for an even smaller effect.

\bibliographystyle{mnras}

\bibliography{ref} 

\bsp	
\label{lastpage}
\end{document}